\pdfoutput=1
\documentclass[10pt,conference,letterpaper]{IEEEtran}

\usepackage[noadjust]{cite}
\usepackage[T1]{fontenc}
\usepackage[utf8]{inputenc}
\usepackage{balance}
\usepackage{url}
\usepackage{todonotes}

\newcommand{\docauthor}{Jonathan Will, Onur Arslan, Jonathan Bader, Dominik Scheinert and Lauritz Thamsen}
\newcommand{\docsubject}{Technische Universit\"at Berlin}
\newcommand{\dockeywords}{Scalable Data Analytics, Distributed Dataflows, Resource Allocation, Machine Learning, Data Reduction}
\newcommand{\doctitle}{Training Data Reduction for Performance Models\\of Data Analytics Jobs in the Cloud}

\PassOptionsToPackage{bookmarks=false}{hyperref}
\usepackage{subfig}
\usepackage[pdftex]{hyperref}
\hypersetup{
    pdfborder={0 0 0},
    pdfauthor=\docauthor,
    pdftitle=\doctitle,
    pdfsubject=\docsubject,
    pdfkeywords=\dockeywords
}

\usepackage{amsmath,amssymb,amsfonts}
\usepackage{algorithmic}
\usepackage{graphicx}
\usepackage{textcomp}
\usepackage{tikz}

\newcommand\copyrighttext{%
  \footnotesize \textcopyright 2021 IEEE. Personal use of this material is permitted.
  Permission from IEEE must be obtained for all other uses, in any current or future
  media, including reprinting/republishing this material for advertising or promotional
  purposes, creating new collective works, for resale or redistribution to servers or
  lists, or reuse of any copyrighted component of this work in other works.
  DOI: \href{https://doi.org/10.1109/BigData52589.2021.9671742}{https://doi.org/10.1109/BigData52589.2021.9671742}}
\newcommand\copyrightnotice{%
\begin{tikzpicture}[remember picture,overlay]
\node[anchor=south,yshift=10pt] at (current page.south) {\fbox{\parbox{\dimexpr\textwidth-\fboxsep-\fboxrule\relax}{\copyrighttext}}};
\end{tikzpicture}%
}

\def\BibTeX{{\rm B\kern-.05em{\sc i\kern-.025em b}\kern-.08em T\kern-.1667em\lower.7ex\hbox{E}\kern-.125emX}}

\begin{document}

\title{\doctitle}

\author{
\IEEEauthorblockN{Jonathan Will\IEEEauthorrefmark{1}, Onur Arslan\IEEEauthorrefmark{1}, Jonathan Bader\IEEEauthorrefmark{1}, Dominik Scheinert\IEEEauthorrefmark{1}, and Lauritz Thamsen\IEEEauthorrefmark{3}\IEEEauthorrefmark{1}}
\IEEEauthorblockA{\IEEEauthorrefmark{1}Technische Universit{\"a}t Berlin, Germany, \{firstname.lastname\}@tu-berlin.de}
\IEEEauthorblockA{\IEEEauthorrefmark{3}Humboldt-Universit{\"a}t zu Berlin, Germany, lauritz.thamsen@hu-berlin.de}
}

\maketitle
\copyrightnotice


\begin{abstract}
Distributed dataflow systems like Apache Flink and Apache Spark simplify processing large amounts of data on clusters in a data-parallel manner.
However, choosing suitable cluster resources for distributed dataflow jobs in both type and number is difficult, especially for users who do not have access to previous performance metrics.
One approach to overcoming this issue is to have users share runtime metrics to train context-aware performance models that help find a suitable configuration for the job at hand.
A problem when sharing runtime data instead of trained models or model parameters is that the data size can grow substantially over time.

This paper examines several clustering techniques to minimize training data size while keeping the associated performance models accurate.
Our results indicate that efficiency gains in data transfer, storage, and model training can be achieved through training data reduction.
In the evaluation of our solution on a dataset of runtime data from 930 unique distributed dataflow jobs, we observed that, on average, a 75\% data reduction only increases prediction errors by one percentage point.

\end{abstract}


\section{Introduction}\label{sec:introduction}
Distributed dataflow systems like Apache Spark~\cite{spark} and Apache Flink~\cite{flink} simplify the development of data processing applications.
Users can write programs with fault tolerance and parallelization being provided by the framework.
The resources to be used to execute such jobs on large datasets can either be dedicated on-premises clusters or cloud resources, which is increasingly common.
The node types and scaleouts should be chosen wisely in order to avoid bottlenecks on the one hand and low utilization on the other hand.
However, doing this is not straightforward, even for experts~\cite{perforator,Lama_AROMA_2012}.

There are many attempts to automatically configure cluster resources for large data analytics~\cite{ernest,cherrypick,bell,will2021c3o,scheinert2021bellamy,scheinert2021enel}.
However, all of them either rely on performance metrics from previous executions or on profiling runs.
Nevertheless, it is expensive to perform extensive profiling of a job, especially if the job is not recurring very frequently, if at all.
There are attempts to learn from gathered performance metrics to increase the amount of possible training data points, even if they stem from different execution contexts, e.g., different cluster infrastructure, input data characteristics, or algorithm parameters~\cite{will2021c3o,scheinert2021bellamy}.
This allows for different users to collaborate on shared runtime datasets and train context-aware performance models~\cite{will2020towards,will2021c3o}.

Compared to traditional collaborative machine learning approaches, sharing just the training data brings several advantages.
First, no complex coordination with a centralized instance or peers is needed to retrain the models continuously.
The models can also be shared, improved, and extended by collaborators who have access to that training data~\cite{will2020towards}.
The second considerable issue is privacy.
While there are attempts at preserving data privacy in federated learning~\cite{yang2019federated,wittkopp2020decentralized}, researchers repeatedly find vulnerabilities and manage to reconstruct the original data~\cite{fredrikson2015model,shokri2017membership}.
These considerations were taken into account in our previous work, C3O, where users have full control over their data as they make contributions to the public training data repository on an entirely voluntary basis.
The models can then be trained with both globally available public data and locally generated private data.

A remaining issue of sharing training data is that the training data can grow infinitely large.
Reducing training data while retaining model accuracy has already been successfully attempted in the context of training classification models.
Here, clustering to reduce training data is an established preprocessing step for computationally intensive classification tasks.
It has been shown that for many classification tasks, most training data can be omitted while preserving nearly the same accuracy~\cite{wang2014training,sutton2012introduction,el2015comparative}.
While we are facing a regression problem, in our domain, having very dense data is generally less beneficial than having the data spread far throughout the data space in each possible dimension.
This is because the effect of many individual runtime influencing factors of distributed dataflow jobs can be interpolated and sometimes even extrapolated~\cite{will2020towards,bell,ernest}.

This paper examines several clustering techniques to minimize data size while keeping the associated regression models of runtimes accurate.
We show that by using clustering techniques to filter out similar data points, we can compress a training dataset while still maintaining adequate model accuracy.\\
\vspace{-2mm}
\\
\vspace{0mm}
\emph{Contributions}. The contributions of this paper are:
\vspace{0mm}
\begin{itemize}
    \item A discussion about requirements and challenges regarding efficiency aspects in sharing training data for runtime models of distributed data processing in the cloud
    \item An idea to reduce training data for runtime models of distributed data processing in the cloud with the help of clustering methods
    \item An evaluation of our idea on a dataset with 930 unique Spark job executions\\
\end{itemize}
\vspace{-6mm}

\section{Related Work}\label{sec:related_work}
In this section, we first give some background on distributed dataflow systems.
Next, we explain how context-aware performance models can be used to choose good cluster resources for executing distributed dataflow jobs.
Lastly, we mention approaches to reduce training data for machine learning models while preserving adequate accuracy.

\subsection{Distributed Dataflow Systems}

Distributed dataflows are graphs of interconnected data-parallel operators that execute user-defined functions on a set of shared-nothing cluster nodes.
By using high-level programming abstractions, users can create data-parallel programs without worrying about implementing parallelization themselves.
The system translates the user's sequential program into a directed graph of data-parallel operators and eventually into an optimized execution plan.
Such systems also manage error handling.
Failed operations are automatically repeated, and faulty nodes are replaced with intact nodes.
Prominent examples of such systems are Hadoop MapReduce~\cite{mapreduce}, Apache Spark~\cite{spark}, and Apache Flink~\cite{flink}, with the latter two belonging to a newer generation, focusing on in-memory computation.

For large datasets, it becomes necessary to use clusters of multiple machines as opposed to using just a single machine~\cite{bader2021tarema}.
Some researchers and large corporations have access to bare-metal clusters, whereas everyone, including small start-ups, can use public cloud services.
On public cloud services such as Amazon Web Services\footnote{\href{https://aws.amazon.com}{aws.amazon.com, accessed October 2021}},
the nodes can have different memory, CPU, IO, and network capabilities, resulting in over 100 options to choose from\footnote{\href{https://aws.amazon.com/ec2/instance-types/}{aws.amazon.com/ec2/instance-types/, accessed October 2021}}.
The node type along with the scale-out of the cluster needs to be chosen wisely to avoid bottlenecks on the one hand and low resource utilization on the other hand.
Many existing approaches iteratively search for suitable cluster configurations~\cite{cherrypick, hsu2018micky, hsu2018arrow, hsu2018scout, fekry2020tuneful}.
Several other approaches~\cite{ernest, scheinert2021bellamy, shah2019quick, bell, perforator, will2021c3o, scheinert2021potential} build runtime models, which are then used to evaluate possible configurations.
The following subsection will explain the possibilities for managing and sharing training data for these runtime models.

\subsection{Cross-context Performance Data Sharing and Modeling}

Users of distributed dataflow systems typically have some expectations regarding the runtime and cost of their data analytics jobs, especially when large datasets are involved.
To choose a suitable configuration of cluster resources for a distributed dataflow job, one can use performance models that predict the runtime for clusters consisting of different types of nodes and different scale-outs, based on runtime metrics from previous executions.
In case the job is recurring within an organization, there might already be such data available, otherwise it can be generated in dedicated profiling runs.
Performance gains in future executions will then offset the cost of such profiling runs~\cite{ernest,bell}.

In other cases, one might make use of data from similar executions that have taken place in different contexts, e.g., on different hardware or with different job parameters.
There has been work on performance models that consider these different execution contexts of distributed dataflow jobs~\cite{scheinert2021bellamy,will2020towards,will2021c3o}.
Data shared in this way can be used to train context-aware performance models.
Within one data analytics algorithm, the jobs vary in inputs and cluster setup.
This means that, on the one hand, there are differences in key dataset characteristics and algorithm parameters, and on the other hand, the jobs were executed on different node types and scale-outs.
With C3O~\cite{will2021c3o}, we introduced a system for sharing jobs along with context-aware runtime models and runtime data among users from different organizations, attempting to mitigate the cold-start problem.
Another approach, Bellamy~\cite{scheinert2021bellamy}, pre-trains models on job performance data and uses transfer learning to adapt the models once there is a change to the context.
Bellamy assumes organization-internal performance data sharing.

\subsection{Training Data Reduction}

In some domains, for example IoT sensor networks, new data gets generated rapidly and in large amounts~\cite{geldenhuys2021dependable}.
However, some machine learning techniques struggle with increasing data sizes.
A concrete example of this is the support vector machine~\cite{wang2014training}.
Here, computational complexity grows quadratically with the training data size.
It is thus vital to keep the amount of training data as low as possible while still maintaining model accuracy to the degree that is appropriate for the task at hand.

Reducing training data to make machine learning more efficient can be done in two ways.
First, one can reduce the dimensions of each data point and remove less relevant features.
In the context of performance prediction for distributed data analytics, we assume that all features are carefully selected and, therefore, relevant.
Second, one can reduce the number of observations, which will be the focus of this paper.
Clustering to reduce training data has thus far existed as a preprocessing step for computationally intensive classification tasks~\cite{wang2014training,sutton2012introduction,el2015comparative}.
Here, similar training data points are grouped, and only one representative of the group remains in the reduced dataset.
It has been found that in these classification tasks, most training data can be omitted while reaching nearly the same accuracy.
This is due to classification models learning the location of the borders between classes of data points without necessarily needing to consider each data point individually.

\section{The C3O System}\label{sec:idea}\label{sec:idea}
\begin{figure}[!htb]
    \centering
    \includegraphics[width=\columnwidth, keepaspectratio]{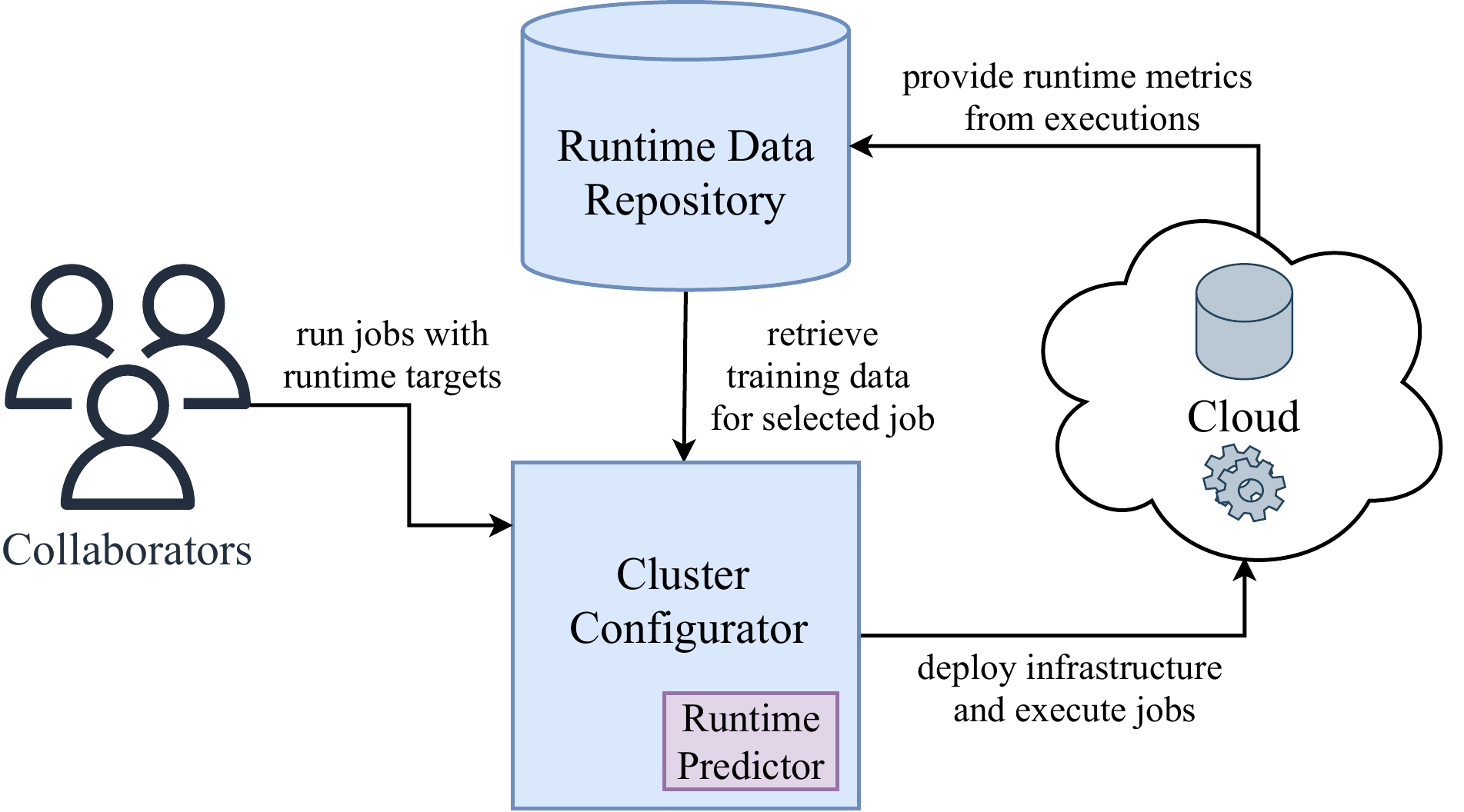}
    \caption{Overview of the C3O system for collaborative cluster configuration optimization for distributed data processing in the cloud (adjusted from~\cite{will2020towards})}\label{fig:c3o}
\end{figure}

C3O is a collaborative system for optimizing data processing cluster configurations in public clouds based on shared historical runtime data~\cite{will2021c3o, will2020towards}.
The shared data is utilized for predicting the runtimes of data processing jobs on different possible cluster configurations, using specialized regression models.
These models need to take the diverse execution contexts of different users into account, e.g., runtime-influencing job parameters and input dataset characteristics.
An overview of the idea behind C3O is given in Figure~\ref{fig:c3o}.

Through the availability of globally shared data, users get to select suitable cluster configurations without the need for previous local executions of the same job.
For one-off jobs, this data availability already enables performance modeling and choosing a suitable cluster configuration accordingly.
For recurring jobs, a user can continuously add his or her own locally generated runtime data to the training dataset.
Upon retraining the model with this new data, possibly in addition to the existing training data, the performance can then be modeled much more accurately for the user's given context.
This is one of the reasons why the approach is centered around sharing training data and models instead of sharing pre-trained models.
Other reasons are considerations of data privacy and data autonomy, and simplified coordination compared to classical collaborative machine learning approaches.

However, an issue that arises when sharing training data is the ever-increasing size of the dataset.
So, if we can reduce the training data while keeping most of the information, C3O would not only benefit from reduced data storage and data transfer costs, but also quicker model training times.

\section{Approach for Training Data Reduction}\label{sec:idea}\label{sec:idea}
In this paper, we examine the feasibility and the possible benefit of reducing the size of collaboratively shared training data for performance models of data analytics jobs.
This section presents the idea in the context of C3O.

\subsection{Training Data Reduction Process}

To collectively decrease the cost of data transfer, data storage and model training, our goal is to limit the amount of training data while retaining the essential information needed to train accurate models.

This results in a trade-off between accuracy and data size.
According to our vision, users will either be able to choose between different pairs of dataset sizes and estimated resulting prediction error rate according to their individual needs, or the system will automatically make this decision.

\begin{figure}[!htb]
    \centering
    \includegraphics[width=\columnwidth, keepaspectratio]{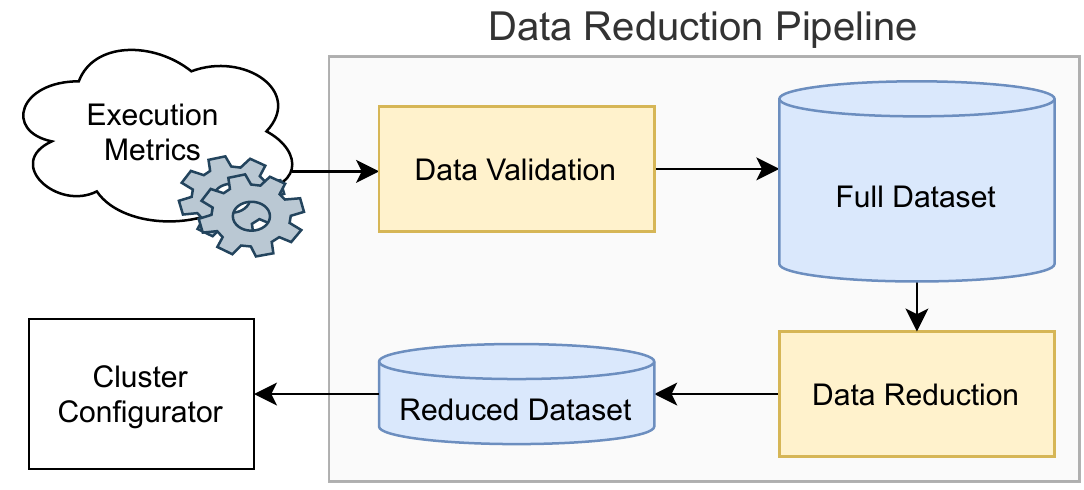}
    \caption{The data reduction pipeline and its interaction with the remaining C3O system}\label{fig:pipeline}
\end{figure}

Our solution takes the form of a pipeline, which is depicted in Figure~\ref{fig:pipeline}.\\
After users submit their newly generated runtime metrics, the new training data is validated for plausibility and quality.
This is a measure to prevent malicious interference with the sharing system.
Therefore, the models are first trained with the new training data, and the system detects if the additional training data causes the models to perform significantly worse.
In this case, the contribution shall be rejected or at least deferred until its addition to the dataset would lead to performance gains.

After validation, the data is admitted into the full training dataset.
Once a significant amount of new data has been received by the system, the reduced training dataset is newly generated and is available to be downloaded by the cluster configurators of the individual users.
The reduced dataset is based on the original training dataset and gained by employing clustering techniques, details of which will be explained in the following subsection.
\subsection{Clustering}

In the particular domain that we are examining, we do have information about the nature of the data.
Here, to achieve accurate prediction results, it is less important to have very dense data.
Instead, having the data spread through every dimension of that space is more beneficial.
Since the individual runtime influencing factors of distributed dataflow jobs are typically straightforward to model with basic functions, this type of data is often simple to interpolate and sometimes even extrapolate~\cite{will2020towards,ernest}.
Moreover, in this domain, it is expected that many data points exist that are very close to each other, due to many data analytics jobs being recurring, especially within a given organization~\cite{jyothi2016morpheus,agarwahl2012reoptimizing}.

Therefore, our strategy is to first remove full duplicates and then to rid the dataset of data points that are very similar to each other with the use of simple, easily reproducible clustering techniques, namely K-Means, K-Medoids, and DBSCAN\@.
We did this with varying amounts of resulting clusters and trained the runtime models on the centroids of the clusters.
To determine the quality of the reduced dataset, we use its data points to train the prediction model and compare the accuracy to the one reached by using the original training dataset.
That way, the trade-off between accuracy and data size can be calibrated.

\section{Evaluation}\label{sec:evaluation}

\begin{figure*}[!htb]
\subfloat[K-Means clustering\label{sfig:kmeans}]{%
  \includegraphics[width=.37\linewidth]{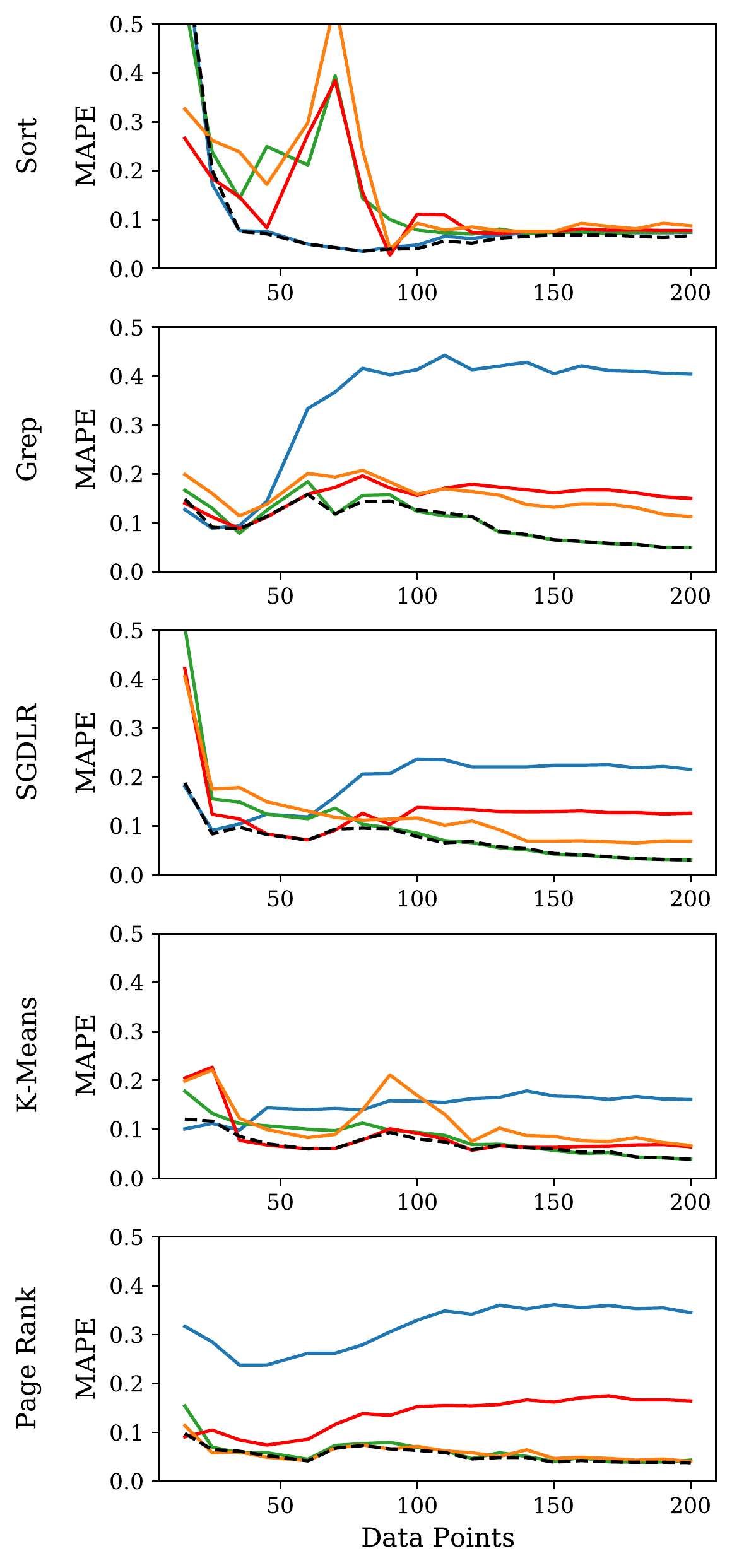}%
}\hfill
    \subfloat[K-Medoids clustering\label{sfig:kmedoids}]{%
  \includegraphics[width=.31\linewidth]{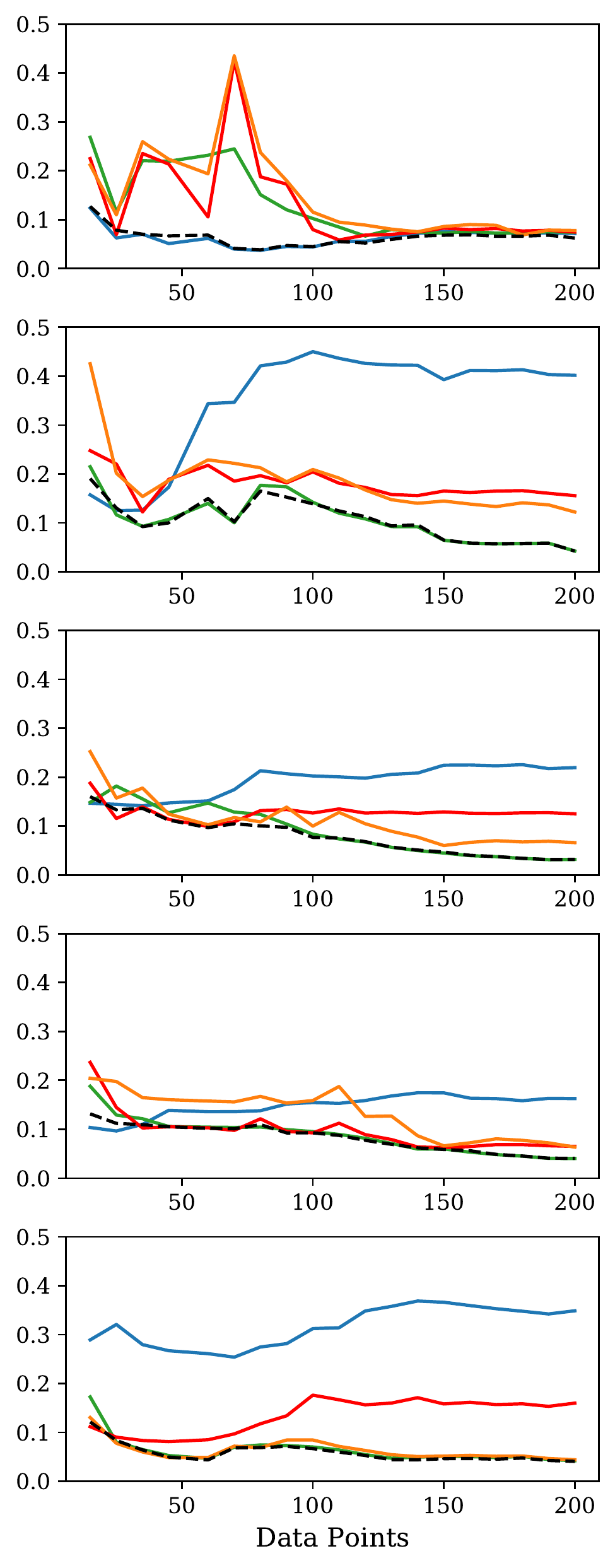}%
}\hfill
    \subfloat[DBSCAN clustering\label{sfig:dbscan}]{%
  \includegraphics[width=.31\linewidth]{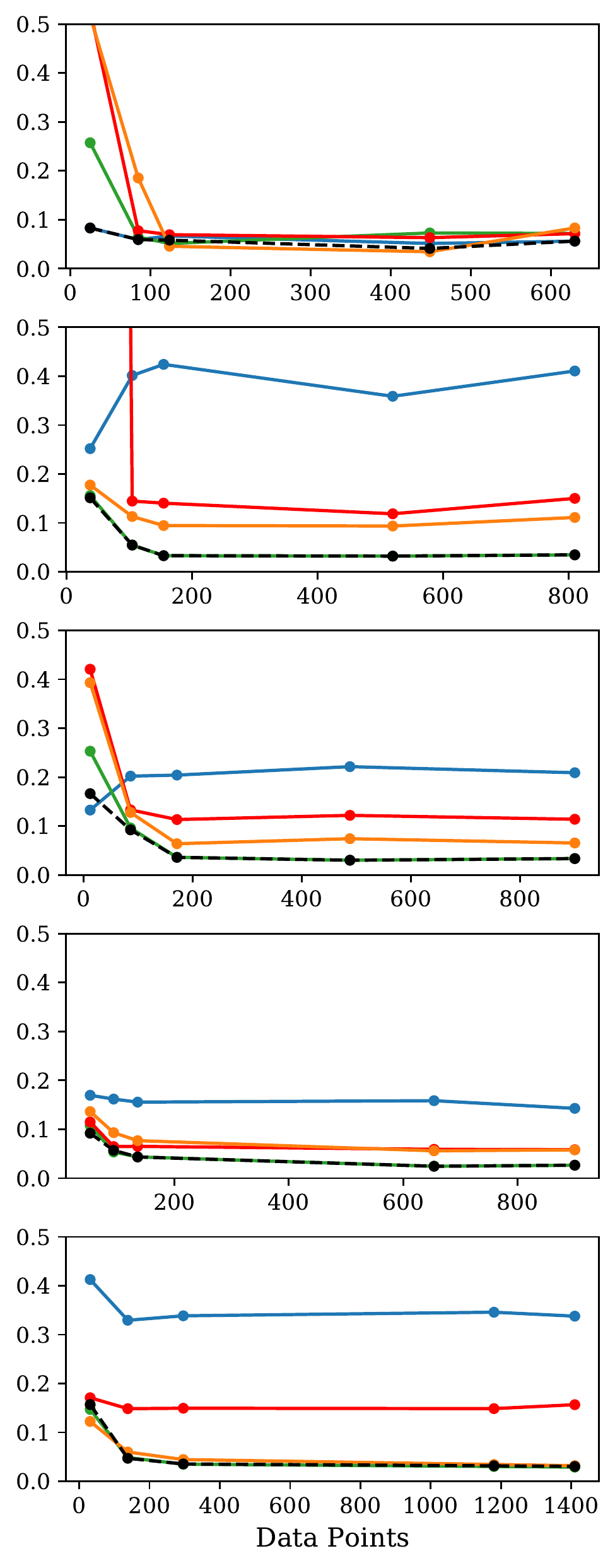}%
}\hfill\centering
    \vspace{-2.2mm}
    \subfloat{\includegraphics[width=.5\textwidth]{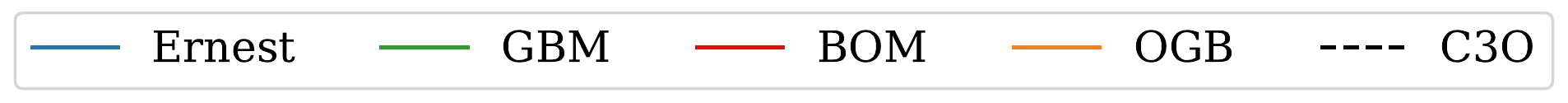}}
    \caption{Mean average percentage error (MAPE) for runtime predictions at different data availabilities. The training data for the models in each plot was generated through (a) K-Means clustering or (b) K-Medoids clustering or (c) DBSCAN clustering.}\label{fig:all}
\end{figure*}

We evaluated our data reduction techniques on the specific use case of runtime estimation for distributed dataflow jobs.
Here, we attempted to reduce the training dataset size significantly, while retaining adequate model accuracy.
The particular metrics that we evaluated in relation to the size of the training data were the runtime prediction error and the training times of the runtime models.

\subsection{Experimental Setup}

For our experiments, we used a dataset of 930 unique Spark jobs across five different data analytics algorithms, details of which can be seen in the associated publication~\cite{will2020towards}.
This dataset includes jobs for Sort, Grep, linear regression with Stochastic Gradient Descent (SGD), K-Means, and Page Rank.
Within one data analytics algorithm, the jobs vary in inputs and cluster setup.
Thus, on the one hand, there are differences in key dataset characteristics and algorithm parameters.
On the other hand, the jobs were executed on different node types and scale-outs.

The runtime predictor used for this evaluation is the C3O runtime predictor, which consists of several runtime models.
These are:
\begin{itemize}
    \item Ernest
    \item Gradient Boosting Model (GBM)
    \item Basic Optimistic Model (BOM)
    \item Optimistic Gradient Boosting (OGB)
\end{itemize}
\vspace{3mm}
The C3O runtime predictor selects one of these individual models according to the given situation based on cross-validation.
The functioning and implementations of all models are described in further detail in the associated publication~\cite{will2021c3o}.

For all clustering algorithms in our experimental evaluations, we used the standard implementations of the \emph{Scikit-Learn} (v. 0.24.2) library for Python~\cite{scikit-learn}.
Each of the experiments was conducted 200 times, and the average results were reported in this section.

For the experiments, we used a personal computer with 16~GB RAM and an i7-9700K CPU at 3.60~GHz.

\subsection{Data Reduction and Model Accuracy}

In the following, we examine the accuracies of the different models within the C3O runtime predictor on reduced training datasets.
The reduction was realized by using three different clustering algorithms.

We varied the amounts of clusters for K-Means and K-Medoids and trained the models on the resulting centroids.
For DBSCAN, on the other hand, we varied the parameter $\epsilon$, which determines the size of a neighborhood.
This, in turn, varied the number of clusters, of which we again considered only the centroids.
In the first steps, this also automatically removed duplicates and near-duplicates that were present in the dataset.

Figure~\ref{fig:all} shows the evaluation of K-Means, K-Medoids and DBSCAN clustering on the datasets of each of the five jobs.
In all instances, we can observe that the individual models react differently to a reduction in available training data.
Also, for all clustering methods, the overall prediction accuracy, represented by C3O, remains largely similar to the fully unreduced scenario.
This is especially the case as long as about a quarter of the amount of data compared to the full training dataset is still available.
We see that different models react differently to a reduction in training data.
For instance, Ernest seems to be less negatively affected than other models like gradient boosting, which might be due to the lower complexity of Ernest.

However, the general observation of more training data leading to more accurate models is visible.
Again, we can also see that different models react differently to a reduction in training data.

\begin{figure}[!h]
    \centering
    \includegraphics[width=\columnwidth, keepaspectratio]{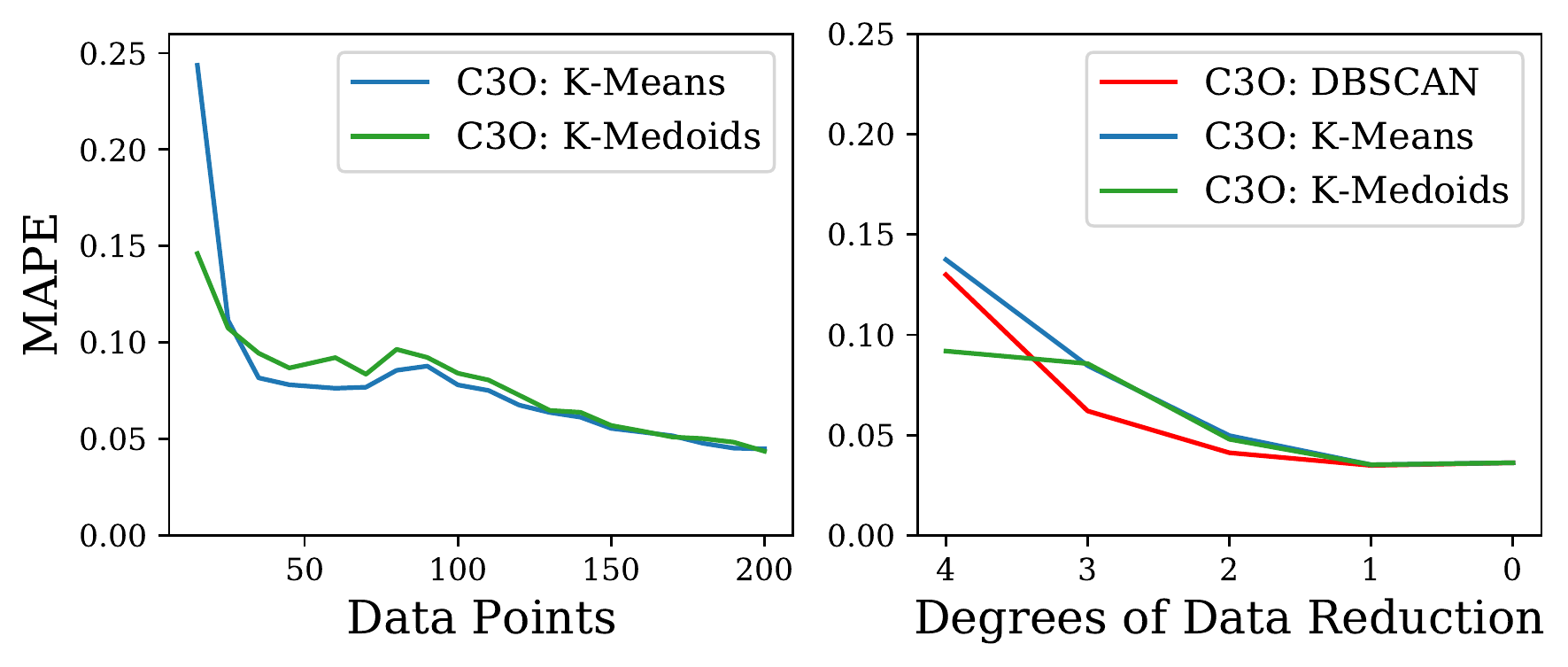}
    \caption{Direct comparison of model accuracy for all data clustering techniques using the same amount of resulting training data points}\label{fig:comparison}
\end{figure}

\pagebreak
Finally, we make a direct comparison of DBSCAN, K-Means, and also K-Medoids across all jobs for the C3O predictor.
For this, we matched the number of centroids in K-Means and K-Medoids to the number of centroids resulting from different clustering granularities in DBSCAN\@.\\
In Figure~\ref{fig:comparison}, we can see that for different degrees of data reduction, DBSCAN and K-Medoids outperform K-Means in terms of retaining model accuracy.
However, the difference in performance between K-Means and K-Medoids only became apparent with high levels of data reduction.
As can also be seen in Figure~\ref{sfig:dbscan}, two degrees of data reduction translates roughly to a reduction by 75\% on average.

\subsection{Data Reduction and Model Training Time}

\begin{figure}[!hbt]
    \includegraphics[width=\columnwidth, keepaspectratio]{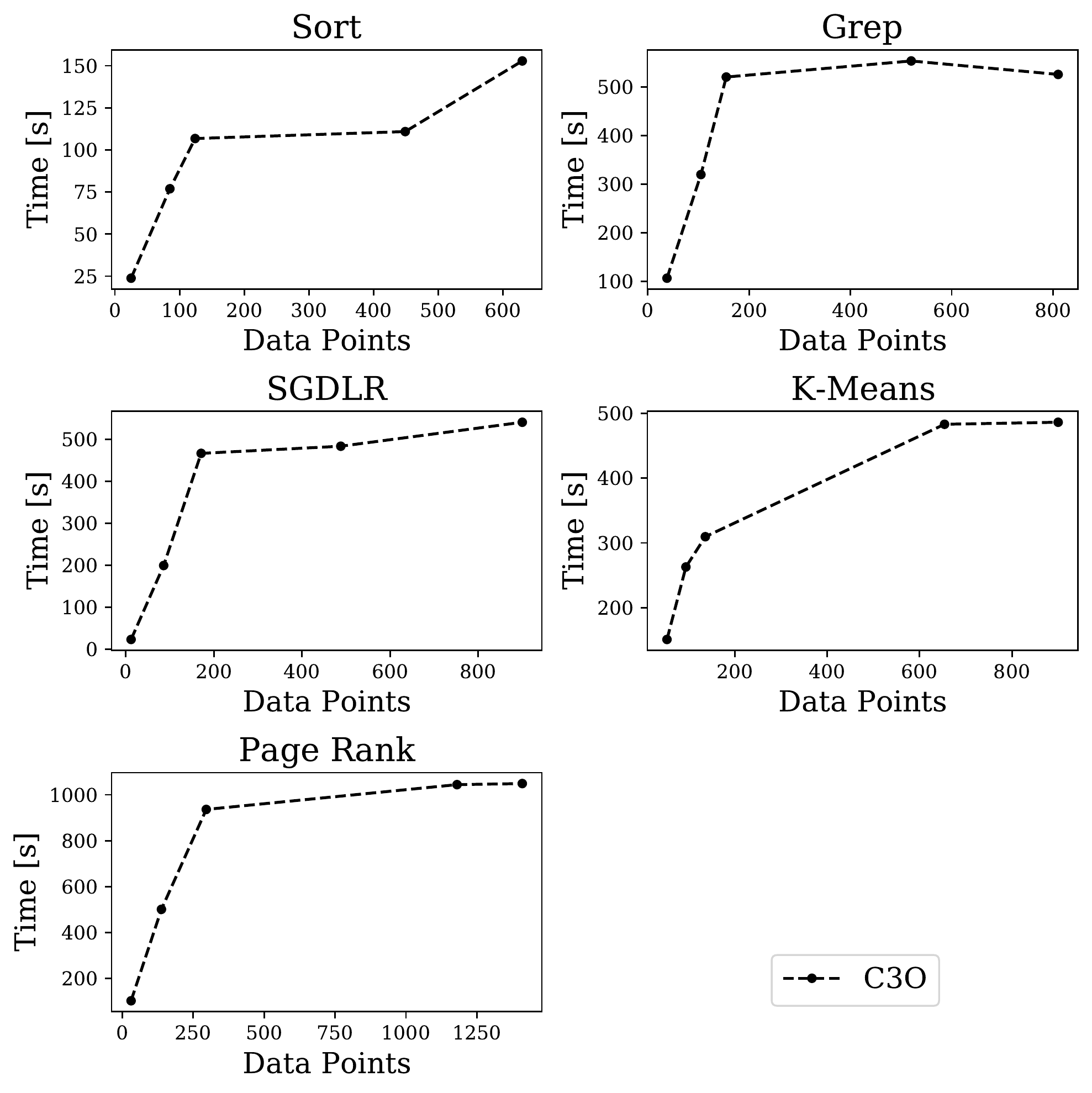}
    \caption{Development of training time of the C3O predictor for varying amounts of training data}\label{fig:time}
\end{figure}

Lastly, we examined the effect of training data size on the training time of the C3O predictor for each of the data analytics algorithms.
Time overhead is one of the challenges of any cluster configuration optimization strategy and is thereby lessened.
In Figure~\ref{fig:time}, we can observe a decrease in model training time on reduced training datasets.
However, in these examined cases, the training time reduction appears to grow slower than linearly in relation to the data reduction.

These results indicate that reduced dataset sizes can reduce not only the cost of data storage and transmission but also reduce the time overhead of downloading and using the training data to train the runtime models.

\pagebreak

\section{Conclusion}\label{sec:conclusion}
We have demonstrated that we can use multiple data compression techniques to reduce training data of runtime models for distributed dataflow jobs significantly without losing prohibitively much information.
In our evaluation, we were able to reach around 75\% reduction in training data, while the mean prediction errors of the models only went up roughly from 4\% to 5\%.
Thus, since the issue of growing datasets can be alleviated, collaborating on shared training data among peers can be a feasible technique for collaborative machine learning in this domain.

Possibilities to expand on this work include finding a simple and efficient way to facilitate the presented data reduction pipeline in a largely decentralized environment.
We believe that it is also worth exploring other domains in which reducing training data for regression models with the help of clustering techniques might be feasible.

\section*{Acknowledgments}

This work has been supported through grants by the German Ministry for Education and Research (BMBF) as BIFOLD (grant 01IS18025A) and the German Research Foundation (DFG) as FONDA (DFG Collaborative Research Center 1404).

\bibliographystyle{IEEEtran}
\balance
\bibliography{./references}

\end{document}